\documentclass{iopconfser}
\usepackage{notoccite}
\usepackage[num]{abntex2cite}				
\usepackage[num]{abntex2cite}
\usepackage[utf8]{inputenc}
\usepackage[utf8]{inputenc}
\usepackage[utf8]{inputenc}
\usepackage[english]{babel}
\usepackage{amsfonts}
\usepackage{amssymb}
\usepackage{verbatim}
\usepackage{xcolor}
\usepackage[T1]{fontenc}
\usepackage{graphicx}
\usepackage{physics}
\usepackage{siunitx}
\usepackage{cite}
\usepackage{makeidx}
\usepackage{helvet}
\usepackage{graphicx}
\usepackage{sectsty}
\usepackage{mdwlist}
\usepackage{tocbibind}
\usepackage{mathtools}
\usepackage{indentfirst}
\newcommand{\bea}   {\begin{eqnarray}}
\newcommand{\eea}   {\end{eqnarray}}
\newcommand{\cc}{”}
\begin{document}

\title{On the Classification of the Lévy-Leblond Spinors}

\author{Luiza Miranda$^{(1)}$, Isaque P. de Freitas$^{(2)}$ and Francesco Toppan$^{(3)}$}

\affil{$~^{(1,2,3)}$COTEO, Centro Brasileiro de Pesquisas Físicas, Rio de Janeiro (RJ), Brazil.}

\email{$~^{(1)}$luiza16@cbpf.br,~$~^{(2)}$isaquepfreitas@cbpf.br,~$~^{(3)}$toppan@cbpf.br}

\begin{abstract}
The first-order Lévy-Leblond differential equations (LLEs) are the
non-relativistic analogous of the Dirac equation: they are the
“square roots” of the Schrödinger equation in ($1+d$) dimensions and
admit spinor solutions. In this paper  we show how to extend to the Lévy-Leblond spinors the
real/complex/quaternionic classification
of the relativistic spinors (which leads to the notions of Dirac,
Weyl, Majorana, Majorana-Weyl, Quaternionic spinors). Besides the free equations, we also consider the presence of potential terms. Applied to a conformal potential, the simplest $(1+1)$-dimensional LLE induces a new differential realization of the $osp(1|2)$ superalgebra in terms of differential operators depending on the time and space coordinates.
\end{abstract}

\section{Introduction}

The Dirac equation which introduces the notion of  relativistic spinors can be regarded, see \cite{dir}, as a ``square root equation\cc ~of the relativistic Klein-Gordon equation. In a similar spirit, L\'evy-Leblond introduced in 1967 a $1+3$-dimensional equation \cite{lle} which is the square root of the nonrelativistic (invariant under the Galilei group) Schr\"odinger equation.  The equation implies the presence of nonrelativistic L\'evy-Leblond spinors. Quite often the L\'evy-Leblond equation (``LLE\cc ~for short) appears when taking, for the speed of light $c$ of a relativistic system, the $c\rightarrow \infty$ limit. On the other hand, LLE has its own virtue; it can be applied to investigate nonrelativistic systems, as those studied in condensed matter, which do not necessarily arise as a limit of relativistic systems. \par
As a clarification note, L\'evy-Leblond correctly criticized the use of the term ``nonrelativistic\cc ~to refer to theories satisfying the Galilei's relativity principle described in {\it Dialogo sopra i due massimi sistemi del mondo}; despite of that, in this paper we adhere to the standard terminology, the adjective ``relativistic\cc ~being exclusively applied to Einstein's 
special relativity.\par
The theory of relativistic spinors is rather rich. Relativistic equations can be generalized to arbitrary $(s,t)$ spacetimes, not only the $1+3$ Minkowski spacetime. Depending on the spacetime, different solutions can be found.
Besides complex spinors (as the original Dirac spinors), real Majorana spinors can be found, Weyl chiral/antichiral spinors can exist which, in certain spacetimes, can be combined to give Majorana-Weyl spinors. Spinors defined for the division algebra of quaternions produce different types of relativistic equations. The classification of the relativistic spinors is based on the \cite{abs}  classification of the associated Clifford algebras.  Relevant works (for this current paper) which present the classification of relativistic spinors are \cite{{kugtow},{oku},{crt}}.\par
Generalized L\'evy-Leblond equations can be introduced as square roots of Schr\"odinger equations in 
$1+d$-dimensions, where the number $d$ of the space components is an arbitrary integer. The Schr\"odinger equations under consideration can be free, but can also admit potential terms. The theories of LLEs and their associated spinors have not been systematically investigated as their relativistic counterparts (see, e.g., \cite{aaca}).\par
Some words should be added to justify the interest in classifying the theories of nonrelativistic spinors. Obviously, information about relativistic theories should be recovered in the $c\rightarrow \infty$ nonrelativistic theory and the mapping between the relativistic spinors and their nonrelativistic counterparts is an essential tool. Similarly, in recent years, the opposite $c\rightarrow 0$ (so-called ``Carroll\cc ) limit of relativistic theories is widely investigated (the connection between Galilei and Carroll limits is discussed in \cite{fig}).  The Galilei algebra is not semisimple and admits the Bargmann's central extension \cite{bar}; due to that its representation theory is quite nontrivial. It also implies, as shown below in Section  {\bf 5}, a different role for the time and for the space coordinates.  Last but not least, in applications to condensed matter emergent nonrelativistic spinors can arise in theories which do not necessarily result from the $c\rightarrow \infty$ limit of relativistic models.\par
In this paper we show how to extend, to the generalized LLEs, the methods presented in \cite{oku} and \cite{crt} for relativistic spinors. It must be said that LLEs turned out to have surprisingly rich structures. The dynamical symmetry of the free equations possess \cite{{aktt1},{aktt2}} $Z_2\times Z_2$-graded Lie superalgebra structures, fitting the class of superalgebras introduced by Rittenberg-Wyler in \cite{{rw1},{rw2}}. Furthermore it was recently shown \cite{ryan} that, in the presence of a harmonic potential, the $Z_2\times Z_2$-graded Lie superalgebras act as spectrum-generating superalgebras.\par
In this paper we show how to systematically classify the  $1+d$-dimensional L\'evy-Leblond spinors as real (that is, of Majorana-type), complex (Dirac-type), quaternionic, chiral/antichiral (Weyl-type) and in determining their vector dimensions. We proceed at first with the free equations; later we show how to introduce potentials. The specialization to the inverse square potentials of the Conformal mechanics \cite{cal}
implies new differential realizations of the superconformal algebras ($osp(1|2)$ in the simplest case) in terms of differential operators in the time and space coordinates.\par
As a technical tool we introduce the convenient \cite{tove} ``alphabetic presentation\cc ~of Clifford algebras which allows to represent the four $2\times 2$ building-block matrices of \cite{crt} as letters and the generic Clifford algebras gamma-matrices as words in a $4$-letter alphabet.

\section{Fundamental ingredients}

The Clifford algebra $Cl(p,q)$ is the enveloping algebra produced by the $n\times n$ matrices $\gamma_i$ which satisfy, for $i,j= 1,2,\ldots, p+q$, the anticommutation relations
\bea\label{gammamat}
    \{\gamma_i,\gamma_j\}&=&2\eta_{ij}{\mathbb I}_{n\times n},
\eea
where ${\mathbb I}_{n\times n}$ is the Identity matrix and $\eta_{ij}$ is a $(p+q)\times (p+q)$ diagonal matrix with $p$ entries
$+1$ and $q$ entries $-1$. In application to relativistic theories $p$ is the number of space-like coordinates and $q$ is the number of time-like coordinates. The irreducible representations of the $Cl(p,q)$ algebras are recovered \cite{{oku},{crt}} from tensoring four $2\times 2$ real matrices. They can be identified with four letters according to
{\footnotesize{
\bea\label{4letters}
&X= \left(\begin{array}{cccc}
         1&0\\
         0&-1
    \end{array}\right),\quad
Y= \left(\begin{array}{cccc}
         0&1\\
         1&0
    \end{array}\right),\quad
    A= \left(\begin{array}{cccc}
         0&1\\
         -1&0
    \end{array}\right),\quad
    I= \left(\begin{array}{cccc}
         1&0\\
      0&1
    \end{array}\right).&
\eea
}}
By dropping the symbol ``$\otimes$\cc ~in the tensor products expressing the generic $\gamma_i$ matrices entering (\ref{gammamat}), we can associate these gamma matrices with words written in a $4$-letter alphabet, see \cite{tove} for details. The three single-letter words $X,Y,A$ define the three gamma matrices entering the $Cl(2,1)$ Clifford algebra. In the \cite{crt} recursive construction the five $4\times 4$ gamma matrices defining $Cl(3,2)$ can be expressed as the two-letter words $XX, XY, XA, YI, AI$ (the three matrices $XX,XY,YI$ are space-like, while the two matrices $XA,AI$ are time-like). The general construction is presented in \cite{crt}. \par
The length of the words defines the size of the corresponding gamma-matrices. The possibility of having Weyl-type spinors is ensured if all gamma matrices are of block-antidiagonal form. In the alphabetic presentation this requirement implies that all corresponding words should start with an $Y$ or $A$ letter.\par
The ${\mathbb R}/{\mathbb C}/{\mathbb H}$ real/complex/quaternionic structures of the Clifford algebras and of their associated relativistic spinors are obtained, via Schur's lemma, by looking at the most general matrices which, at given $p,q$, commute with all gamma matrices:\\
$\bullet$ in the ${\mathbb R}$ case the most general matrix is proportional to the identity ${\mathbb I}_{n\times n}$,\\
$\bullet$ in the ${\mathbb C}$ case it is given, for $a,b$ reals, by $a{\mathbb I}_{n\times n}+b J$, where $J^2=-{\mathbb I}_{n\times n}$,\\
$\bullet$ in the ${\mathbb H}$ case it is given, for $a,b_1,b_2,b_3$ reals, by $a{\mathbb I}_{n\times n}+\sum_i b_i J_i$, where the three matrices $J_i$ are imaginary quaternions satisfying the equations $J_iJ_j=-\delta_{ij}{\mathbb I}_{n\times n} +\varepsilon_{ijk}J_k$.\\
The three above cases find an adequate description in terms of the alphabetic presentation.\par
~\\
The extension of the classification to the L\'evy-Leblond nonrelativistic spinors requires introducing, in the alphabetic presentation, a ``fifth letter\cc ~denoted by $Q$. It correspond to the $2\times 2$ differential matrix
{\footnotesize{\bea\label{qmatrix}
Q= \left(\begin{array}{cc}
         0&1\\
         i\partial_{t}&0
    \end{array}\right)  &\Rightarrow& Q^2 = i\partial_{t}
    \left(\begin{array}{cc}
         1&0\\
         0&1
    \end{array}\right).
\eea
 }}
Essentially, $Q$ is the square root of the time derivative entering the nonrelativistic Schr\"odinger equation.\par
An important property in the construction of LLEs is the anticommutation of the matrices associated with the letters $Q$ and $X$:
\bea
\{Q,X\}&=&0.
\eea

\section{L\'evy-Leblond free equations and types of spinors}

By setting $\hbar=1$ and $m=\frac{1}{2}$ the free (matrix) Schr\"odinger equation in $1+d$ dimensions reads as
\bea
i\hbar\partial_t\Psi(t,{\vec x})= -\frac{{\vec{\nabla}^2}}{2m} \Psi(t,{\vec x}) &\Rightarrow& i\partial_t \Psi(t,{\vec x}) = -{\vec{\nabla}}^2\Psi(t,{\vec x}),
\eea
where $\Psi(t,{\vec x})$ is an $n$-component vector (for $n=1,2,3,4,\ldots$). Its simplest square root equation is obtained for $d=1$ and $n=2$.  By using the (\ref{4letters},\ref{qmatrix}) notation it is written as
\bea
Q\Psi(t,x)= X\partial_x\Psi(t,x) &\Rightarrow & i\partial_t\Psi=Q^2\Psi = QX\partial_x\Psi = -X\partial_x Q\Psi=  -X^2\partial_x^2\Psi =-\partial_x^2\Psi.
\eea
This construction admits generalizations. The next simplest case is the square root of the $n=4$-components $d=2$ free Schr\"odinger equation with space coordinates $x,y$. By taking an alphabetic basis of the five gamma matrices of the Clifford algebra $Cl(3,2)$, given by $XX, XY, XA, YI, AI$, one can express the square root equation as
\bea
QI\Psi &=& (XX\partial_x+XY\partial_y)\Psi.
\eea
The corresponding $4$-component spinors are real (that is, they are the nonrelativistic analogues of the Majorana spinors). \par
The $4$-component equation 
\bea
QY\Psi (t,x)&=& XY\partial_x\Psi (t,x)
\eea
admits spinors which are both real and (due to the presence of the block-antidiagonal matrix $Y$) chiral; they are the nonrelativistic analogues of the Majorana-Weyl spinors. Due to the chirality constraint its fundamental components, in real counting, are $\frac{1}{2}\times 4 =2$. \par
~\par
The next level of square root equations is obtained for $n=8$-component spinors. The building blocks are the gamma matrices of the $Cl(4,3)$ Clifford algebras. Two equivalent alphabetic presentations are given by the following two sets of seven $3$-letter words.\\ The first set is given by $XXX, XXY, XXA, XYI, XAI, YII, AII$. \\ The second set by $XYX, XYY, XYA, XXI, XAI, YII, AII$.\par
Three inequivalent square root equations are obtained.\\ The first one is
\bea\label{eq1}
QII \Psi &=& (XXX\partial_x+XXY\partial_y+XYI\partial_z)\Psi,
\eea
producing $8$-component Majorana-type spinors for a $1+3$-dimensional free LLE (the space coordinates being $x,y,z$).\\
The second equation is
\bea\label{eq2}
QII \Psi &=& XYI\partial_x\Psi.
\eea
The $IIA$ complex structure (which commutes with both $QYI$ and $XYI$) implies complex (nonrelativistic analogues of Dirac) spinors for an $8$-component $(1+1)$-dimensional free LLE.\\
The third equation is 
\bea\label{eq3}
QYI \Psi &=& (XYX\partial_x+XYY\partial_y)\Psi.
\eea
The presence of the $Y$ letter in the second position implies that it describes nonrelativistic Majorana-Weyl type spinors which, due to chirality, possess $\frac{1}{2}\times 8 = 4$ components. The equation (\ref{eq3}) is an $8$-component $(1+2)$-dimensional  free LLE.\par
~\par
The next level is obtained for $n=16$.  At this level one recovers $5$ inequivalent free L\'evy-Leblond equations. They are:
\\~\\
$\bullet$
the $16$-component ``Majorana-type\cc ~spinor for the $(1+4)$-dimensional free LLE, given by
\bea
QIII\Psi &=& (XXXX\partial_x+XXXY\partial_y+XXYI\partial_z+XYII\partial_w)\Psi;
\eea
$\bullet$
the ``Dirac-type\cc ~complex spinor for the $(1+2)$-dimensional free LLE with $IIIA$ complex structure,
\bea
QIII\Psi &=& (XXYI\partial_x+XYII\partial_y)\Psi;
\eea
$\bullet$
the ``Majorana-Weyl-type\cc ~spinor (with $8$ real components) for the $(1+3)$-dimensional free LLE 
\bea
QYII\Psi &=& (XYXX\partial_x+XYXY\partial_y+XYYI\partial_z)\Psi;
\eea
$\bullet$ the ``Weyl-type\cc ~complex spinor for the $(1+1)$-dimensional free LLE with $IIIA$ complex structure,
\bea
QYII\Psi &=& XYYI\partial_x\Psi;
\eea
$\bullet$ the quaternionic spinor for the $(1+1)$-dimensional free LLE with $IIIA, IIAX, IIAY$ quaternionic structure,
\bea
QIII\Psi &=& XYII\partial_x\Psi.
\eea
We presented here the inequivalent free LLEs for $n=2,4,8,16$, that is $n=2^k$ for $k=1,2,3,4$. The construction can be easily extended to any $k$. It is based on the properties of the associated $Cl(p,q)$ Clifford algebras. Up to 
$k\leq 4$ we can present the following table illustrating the Majorana-type ($M$), Weyl-type ($W$), Majorana-Weyl-type
($MW$), Dirac-type ($D$), Quaternionic ($H$) nonrelativistic spinors of the free $(1+d)$-dimensional LLEs. We get

\bea
{\textrm{$(2\times 2)$ matrices:}} && M, \qquad (1+1), \qquad {\textrm{$2$ real components}},\nonumber\\
{\textrm{$(4\times 4)$ matrices:}} && M, \qquad (1+2), \qquad {\textrm{$4$ real components}},\nonumber\\
{\textrm{$(4\times 4)$ matrices:}} && MW, \quad~ (1+1), \qquad {\textrm{$4/2=2$ real components}},\nonumber\\
{\textrm{$(8\times 8)$ matrices:}} && M, \qquad (1+3), \qquad {\textrm{$8$ real components}},\nonumber\\
{\textrm{$(8\times 8)$ matrices:}} && MW, \quad~ (1+2), \qquad {\textrm{$8/2=4$ real components}},\nonumber\\
{\textrm{$(8\times 8)$ matrices:}} && D, \qquad~ (1+1), \qquad {\textrm{$4_{\bf C}\equiv 8$ real components}},\nonumber\\
{\textrm{$(16\times 16)$ matrices:}} && M, \qquad (1+4), \qquad {\textrm{$16$ real components}},\nonumber\\
{\textrm{$(16\times 16)$ matrices:}} && MW, \quad~ (1+3), \qquad {\textrm{$16/2=8$ real components}},
\nonumber\\
{\textrm{$(16\times 16)$ matrices:}} && D, \qquad~ (1+2), \qquad {\textrm{$8_{\bf C}\equiv 16$ real components}},
\nonumber\\
{\textrm{$(16\times 16)$ matrices:}} && W, \qquad (1+1), \qquad {\textrm{$4_{\bf C}\equiv 8$ real components}},\nonumber\\
{\textrm{$(16\times 16)$ matrices:}} && H, \qquad (1+1), \qquad {\textrm{$4_{\bf H}\equiv 16$ real components}}.
\eea

\section{The introduction of potential terms}

In this Section we show how our scheme can be enlarged to introduce L\'evy-Leblond equations which are square roots of matrix Schr\"odinger equations in the presence of potential terms. We illustrate the simplest example of the
nonrelativistic Majorana spinors in $1+1$ dimensions. \par
The introduction of potential terms requires larger matrix realizations with respect to  the free case.  While  the free L\'evy-Leblond equation in $1+1$ dimensions can be realized by $2\times 2$ matrices, we need at least $4\times 4$ matrices to include the potentials. The construction goes as follows. The basic equation is
\bea
    QI\Psi&=&XY\partial_{x}\Psi +XAf(x)\Psi,
\eea
where the ``words"  $QI$, $XY$, and $XA$ represent $4\times 4$ matrices; the symbol $\partial_x$ denotes the derivative with respect to the space coordinate $x$ and the function $f(x)$ is the prepotential which allows to reconstruct the potential terms of the matrix Schr\"odinger equation. The column vector spinor $\Psi(x,t)$ has 
$4$ components ($\Psi^T=(\psi_1,\psi_2,\psi_3,\psi_4$)). The following equations are satisfied for its $\psi_i(x,t)$ components with $i=1,2,3,4$:
\bea\label{psi34}
    \psi_{3}& =& \partial_{x}\psi_2+f(x)\psi_2,\nonumber\\
    \psi_{4} &=& \partial_{x}\psi_1-f(x)\psi_1,
\eea
together with the time-dependent equations
\bea\label{psi12}
i\partial_{t}\psi_1&=&-\partial_{x}\psi_4-f(x)\psi_4,\nonumber\\
i\partial_{t}\psi_2&=&-\partial_{x}\psi_3+f(x)\psi_3.
\eea
The (\ref{psi34}) algebraic equations for $\psi_3,\psi_4$ imply two independent Schr\"odinger equations for $\psi_1,\psi_2$, given by
\bea
i\partial_{t}\psi_1&=&-\partial^2_{x}\psi_1+\left(f^2(x)+f'(x)\right)\psi_1,\nonumber\\
i\partial_{t}\psi_2&=&-\partial^2_{x}\psi_2+\left(f^2(x)-f'(x)\right)\psi_2,
\eea
where $f'(x)=\frac{d}{dx}f(x)$ and 
\bea
V_\pm (x) &=& f^2(x)\pm f'(x)
\eea 
are the respective potentials.\par
By applying the $i\partial_t$ derivative to $\psi_3,\psi_4$ entering (\ref{psi34}) and taking into account the (\ref{psi12}) equations, we obtain that the $\psi_3,\psi_4$ components satisfy the Schr\"odinger equations 
\bea
i\partial_{t}\psi_3&=&-\partial^2_{x}\psi_3+V_+(x)\psi_3,\nonumber\\
i\partial_{t}\psi_4&=&-\partial^2_{x}\psi_4+V_-(x)\psi_4.
\eea
The extension of the procedure to introduce potentials for nonrelativistic spinors in ($1+d)$-dimensions, complex structures, etc., is rather straightforward; it is based on the previously discussed construction of the free equations. \par
The specialization to certain types of potentials, like the  inverse square potential proportional to $\frac{1}{x^2}$,
produces interesting differential realizations of the superalgebras associated with the L\'evy-Leblond square root of the matrix conformal mechanics. The simplest example is illustrated in the next Section.

\section{The $osp(1|2)$ superalgebra induced by a conformal potential}

In $1+1$ dimensions, the L\'evy-Leblond ``square root” of a matrix conformal mechanics with potential term proportional to $1/{x^2}$ is given by the $4\times 4$ matrix differential equation
\bea\label{confll}
\Omega \Psi (x,t) &=& 0,
\eea
where the $\Psi(x,t)$ L\'evy-Leblond spinor is a $4$-component column vector and the matrix differential operator $\Omega$ belongs to a $5$-generator $osp(1|2)$ superalgebra. The explicit form of the five $osp(1|2)$ generators is given in terms of the $2\times 2$ matrices $I,X,Y,A$ and $Q$ introduced in (\ref{4letters}) and (\ref{qmatrix}), plus two extra $2\times 2$ auxiliary matrices $\Lambda$ and $R$ which depend on an (arbitrary) real scaling parameter $\lambda$:
\bea
\Lambda =   \left(\begin{array}{cc} \lambda &0\\ 0&\lambda+\frac{1}{2}\end{array}\right),&& R = \left(\begin{array}{cc} 0 &0\\ \lambda&0\end{array}\right).
\eea
The consistent scaling assignments of the time/space coordinates $t,x$, of their derivatives  and of the auxiliary $2\times 2$ matrices $Q,\Lambda,R$ are:
\bea
&[t]=-1, \quad [\partial_t]=+1,\quad [x]=-\frac{1}{2},\quad [\partial_x]=+\frac{1}{2},\quad [Q] = +\frac{1}{2},\quad [\Lambda] = 0, \quad [R] = -\frac{1}{2}.&
\eea
The five $osp(1|2)$ generators (denoted as $H,\Omega, D, {\Xi}, K$) are given by
\bea
H &=& I\otimes I\cdot(i\partial_t+\partial_x^2-\frac{g^2}{x^2}) + I\otimes X\cdot \frac{g}{x^2},\nonumber\\
\Omega &=& Q\otimes I -X\otimes Y\cdot\partial_x- X\otimes A\cdot \frac{g}{x},\nonumber\\
D &=& I\otimes I\cdot (\frac{1}{4} +\frac{1}{2}x\partial_x+t\partial_t) +\Lambda\otimes I,\nonumber\\
\Xi &=& Q\otimes I\cdot (-it) - X\otimes Y\cdot\frac{x}{2} + R\otimes I,\nonumber\\
K&=& I\otimes I\cdot(-it^2\partial_t+\frac{x^2}{4})-\Lambda\otimes I\cdot 2it,
\eea
where $g$ is a dimensionless ($[g]=0$) coupling constant.\par
The scaling dimensions of the above five generators are
\bea
&[H]=+1,\quad [\Omega]=+\frac{1}{2},\quad [D]=0,\quad [\Xi]=-\frac{1}{2},\quad [K]= -1.&
\eea
The three generators $H,D,K$ are even (bosonic) and close an $sl(2)$ subalgebra having $D$ as the Cartan generator. The two generators $\Omega, \Xi$ are odd (fermionic); they respectively correspond to the positive and negative simple roots of $osp(1|2)$.\par
The closure of the $osp(1|2)$ superalgebra is given by the following (anti)commutators:
\bea\label{osp12}
\relax &[D,H]=-H,\qquad [D,K] = K, \qquad [H,K] =2D,&\nonumber\\
\relax &[D,\Omega ] = -\frac{1}{2} \Omega,\quad ~~~ [D,\Xi]= \frac{1}{2}\Xi,\qquad\qquad\qquad\quad~\nonumber\\
\relax&\qquad\qquad\qquad   ~~ [H,\Omega]=0,\qquad\quad  [K,\Omega] = -\Xi,\qquad [K,\Omega]= - \Xi,\quad [K,\Xi]=0,&\nonumber\\
&\{\Omega,\Omega\}= 2H,\qquad \{\Omega,\Xi\} = 2D,\qquad \{\Xi,\Xi \}= 2 K.&
\eea
Therefore, $\Omega$ is the square root of the $4\times 4$ matrix conformal mechanics Schr\"odinger equation 
\bea\label{confschr}
H\Psi(x,t) &=&0.
\eea
Usually, a Schr\"odinger equation such as (\ref{confschr}) is split into a left part which depends on a first-order time-derivative and a right part which depends on a second-order differential equation in $x$ (the Hamiltonian, which we denote in boldface as ${\bf H}$); 
the (\ref{confschr}) equation can therefore be rewritten as
\bea
I\otimes I\cdot i\partial_t \Psi (x,t)&=& {\bf H} \Psi(x,t) \equiv  I\otimes I (-\partial_x^2 +\frac{g^2}{x^2})-I\otimes X\cdot \frac{g}{x^2}.
\eea
Superconformal algebra realizations of first-order in time ``$D$-module\cc ~differential realizations have been investigated in \cite{kuto}.  For the free Schr\"odinger equation,  a second-order differential realization of $osp(1|2)$ in terms of the space coordinate $x$ appears naturally (see \cite{duality}, which connects the \cite{wig} Wigner's approach to the quantization of the harmonic oscillator with the Niederer's analysis \cite{{nie1},{nie2},{nie3}} of the maximal kinematical invariance group of the Schr\"odinger equation for various potentials).\par
The L\'evy-Leblond equation $\Omega\Psi=0$ induces, for the conformal potential proportional to $\frac{1}{x^2}$, a new
interesting and non-trivial differential realization of $osp(1|2)$ which involves both time and space coordinates.
\par
The construction of the admissible Hilbert spaces for the conformal mechanics, depending on the range of the coupling constant $g$, has been discussed in \cite{{mits},{ftf}} and, in the matrix case, \cite{tova}.\par
An open question should be pointed out. The addition in conformal mechanics of a harmonic potential term to the $\frac{1}{x^2}$ potential
 produces the de Alfaro-Fubini-Furlan deformed oscillator \cite{{dff},{hoto}} having the same conformal algebra as spectrum-generating algebra. The new Hamiltonian is the sum of ${\bf H}$ and of its conformal partner. Concerning the 
(\ref{osp12}) realization, one should note that the sum $H+K$ does not produce a L\'evy-Leblond equation.  It is an open question how to produce a de Alfaro-Fubini-Furlan construction for the LLE.

\section{Conclusions}

This paper outlines the classification of the L\'evy-Leblond nonrelativistic spinors and of their associated equations using tools and methods borrowed from the classification of relativistic spinors. It further illustrates in simple examples how to introduce potential terms to L\'evy-Leblond equations and how to derive the ``square root\cc ~of conformal mechanics with their induced superconformal algebras. This work is part of a research program where
an extended version and systematic presentation of the constructions for real/complex/quaternionic and Weyl-type
L\'evy-Leblond spinors is under preparation.

\par
~\par
~
\\ {\Large{\bf Acknowledgments}}
{}~\par{}~\\
 This work was supported by CNPq (PQ grant 308846/2021-4).

\end{document}